# Broadband THz-TDS with 5.6 mW average power at 540 kHz using organic crystal BNA

Samira Mansourzadeh*[1], Tim Vogel[1], Alan Omar[1], Mostafa Shalaby[2], Mirko Cinchetti[3],

Clara J. Saraceno[1]

[1)]Photonics and Ultrafast Laser Science, Ruhr-University Bochum, 44801 Bochum, Germany

[2)]Swiss Terahertz Research-Zurich, Technopark, Zurich, Switzerland

[3)]Department of Physics, TU Dortmund University, Dortmund, Germany

(E-Mail: *mansourzadeh.samira@Ruhr-Uni-Bochum.de)

**We demonstrate efficient optical rectification in the organic crystal BNA (N-benzyl-2-methyl-4-nitroaniline), driven by a temporally compressed, commercially available industrial Yb-laser system operating at 540 kHz repetition rate. Our THz source reaches 5.6 mW of THz average power driven by 4.7 W, 45 fs pulses and the resulting THz-TDS combines a very broad bandwidth of 7.5 THz and a high dynamic range of 75 dB (in a measurement time of 70 s). The conversion efficiency at maximum THz power is 0.12%. To the best of our knowledge, this is the highest THz power so far demonstrated with BNA, achieved at a high repetition rate, and enabling to demonstrate a unique combination of bandwidth and dynamic range for THz-spectroscopy applications.**

## I. Introduction

Few-cycle terahertz (THz) sources driven by ultrafast lasers are widely used in THz time domain spectroscopy (THz-TDS) which is a well-established tool for many applications in THz science and technology [1–4]. For applications where ultra-broad THz bandwidths are desired, most commonly in spectroscopy, various techniques are available to generate broadband THz emission such as two-color plasma filaments, optical rectification and spintronic THz emitters. Among these techniques, collinear optical rectification (OR) in nonlinear crystals with second order nonlinearity is one of the simplest and most commonly used methods, albeit well-known difficulties in achieving very broad bandwidths due to phonon absorption peaks in typically used inorganic materials such as Gallium phosphide (GaP) or Zinc telluride (ZnTe). In this respect, organic crystals such as DAST, DSTMS, OH1, HMQ-TMS and BNA provide a promising platform to overcome this drawback. Due to the intrinsically lower dispersion of the refractive index in organic nonlinear crystals, they are collinearly phase-matched in a much broader THz bandwidth compared to inorganic crystals [5,6]. For example, Puc et al [7] detected signal bandwidth more than 20 THz using DSTMS pumped by 38 fs laser pulses. In addition, typically high nonlinear coefficient results in very high optical to THz conversion efficiencies, reaching up to the percent level. Using DSTMS, Vicario et al [8] achieved conversion efficiency of 3% which led to ultra-high THz pulse energy of 0.9 mJ at repetition rate of 10 Hz. For the organic crystal OH1, a conversion efficiency of 3.2% was demonstrated, again at low repetition rate of 10 Hz [9]. By pumping a DAST crystal with mid-IR pulses, Gollner et al. achieved a record value of 6% conversion efficiency and 1 mW of THz average power at repetition rate of 20 Hz [10]. BNA is also very-well suited for intense THz generation [11] and has already shown to operate efficiently with Ytterbium-based lasers [12]. However, due to limited crystal quality and poor thermal properties (compared to inorganic crystals) most results using these crystals were so far limited to operation with low repetition rate < 1 kHz pump lasers with high pulse energy and low average power. Only very recently, these crystals have started to be investigated for high repetition rate regimes: HMQ-TMS and BNA recently showed first promising results with THz average power of around 1 mW at >10 MHz repetition rates [13,14]. In these results, the thermal limitations can be circumvented by operating in burst mode, leaving enough time between bursts for the crystal to cool down [14].

However, even if thermal effects can be controlled, the optical rectification process is challenging to drive efficiently at MHz repetition rates. This is due to comparatively low pulse energies being available, unless prohibitively high average powers are used, which are hardly commercially available. As a consequence, the demonstrated multi-MHz repetition rate sources still operate with moderate conversion efficiency and yield rather moderate THz fields, which limits many applications in spectroscopy. Furthermore, repetition rates of tens of MHz could for some systems be too fast to allow the samples to relax from pulse to pulse, which is a prerequisite for pump-probe experiments. In this respect, repetition rates of hundreds of kHz offer an attractive middle ground for high-dynamic range and intense THz generation using well-established industrial-grade lasers. So far, only one recent result has reported optical rectification in this excitation regime using GaP and GaSe as emitters [15], but no reports were made using organic crystals.

In this letter, we demonstrate a THz-TDS based on collinear OR in the organic crystal BNA (Swiss Terahertz GmbH) at room temperature, driven by a commercial laser system delivering up to 47 W of average power, operating at a 540 kHz repetition rate. The pulses from the driving laser are temporally compressed from 240 fs down to 45 fs using a home-built Herriott-type multi-pass cell (MPC) compressor. Using 4.7 W of driving power on the crystal and a duty cycle of 50%, we reach 5.6 mW of THz average power with a broad bandwidth extending up to 7.5 THz at a high dynamic range of 75 dB, with a conversion efficiency of 0.12%. To the best of our knowledge, this is the highest average power THz source achieved so far in BNA, improving the current state-of-the-art reported in [14] by factor of more than 5 and an order of magnitude higher conversion efficiency. Moreover, it is the first organic crystal-based source operating in the attractive multi-hundred kHz repetition rate regime.

## II. Experimental setup and methods

### A. Laser system and pulse compressor

The experimental setup is shown in Fig. 1. The driving laser is a commercial Ytterbium-doped laser (Carbide, Light Conversion) with a central wavelength of 1035 nm, a maximum average power of 50 W and a pulse duration of 240 fs. The current experiment is performed at a repetition rate of 540 kHz; however, the laser could potentially be tuned in repetition rate between 0.1 MHz to 1 MHz. In order to both increase the efficiency of THz generation in BNA and generate broader bandwidth THz radiation, an external compressor is beneficial to shorten the temporal duration of the laser pulses. The Herriott-type MPC, indicated by a dashed-contour box, has been designed and built in-house. It consists of two highly reflective plano-concave mirrors with different radii of curvature (ROC) of 300 mm and 500 mm and a cavity length of 750 mm between these mirrors, providing 13 roundtrips (26 passes) through a 9.5-mm anti reflection coated fused silica (FS) plate, which represents the nonlinear medium where spectral broadening of our laser takes place via self-phase modulation. The plate is carefully positioned along the caustic of the beam to achieve sufficient broadening without spatio-temporal couplings that can degrade the output beam quality. One of the two mirrors has a group delay dispersion (GDD) of -350 $fs^2$ per bounce which compensates for the material dispersion of the nonlinear medium. After spectral broadening in the MPC, the beam undergoes 12 reflections on dispersive mirrors with a GDD of -200 $fs^2$ per bounce to remove the chirp from the spectrally broadened pulse and compress it temporally down to a 45-fs pulse duration, which is very close to the Fourier limit of the broadened spectrum.

We characterize the compressed pulses at the aforementioned repetition rate utilizing second harmonic generation frequency resolved optical gating (SHG-FROG). Fig. 2(a) and Fig. 2(b) show the measured and reconstructed SHG-FROG traces, exhibiting a retrieval error of 0.3 % on a 512x512 grid, confirming the good fidelity of the retrieved trace. Fig. 2(c) shows the temporal pulse profile of the retrieved pulses. The achieved peak power at this repetition rate is 1.4 GW, making this a very attractive high-repetition rate laser system for optical rectification, but also for other nonlinear conversion applications. Fig. 2(d) shows the measured and reconstructed spectrum and spectral phase after MPC, and it is compared to the spectrum measured using an optical spectrum analyzer (OSA).

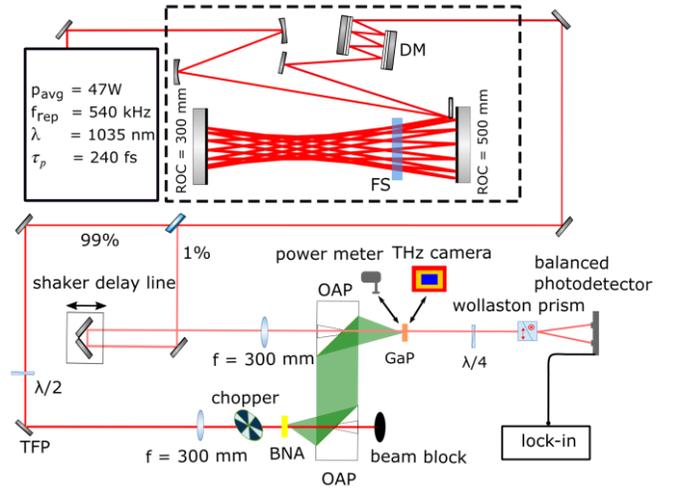

Fig. 1. Full experimental setup consisting of driving laser, MPC and THz-TDS setup. TFP: Thin film polarizer, OAP: off axis parabolic mirror, ROC: radius of curvature, GDD: group delay dispersion, DM: dispersive mirrors, FS: fused silica.

### B. THz -TDS

After the compressor, the laser beam is guided towards the THz-TDS setup and is split in two parts: 99% of the incoming laser power is used to pump a 0.65 mm thick BNA crystal glued on a sapphire substrate to generate THz radiation by OR and 1% is used as probe beam for electro-optic sampling. The $1/e^2$ diameter of the laser beam on the position of BNA is 1.6 mm, generated by a focusing lens with a focal length of 300 mm, placed before the crystal. As we show in our recent exploration at higher repetition rates [14], it is critical for operation of such crystals at high repetition rate to use a chopper wheel to approach as much as possible the thermal relaxation time of the crystal during the off-time of a burst. For this experiment, we use an optical chopper wheel with a duty cycle of 50% which was the only one available at the time of this experiment. Note that the duty cycle could be reduced even further to reach possibly higher efficiencies and/or use the full power of the laser system. The pump power is adjusted on the BNA crystal using a combination of a thin film polarizer (TFP) and a waveplate ($\lambda/2$). The generated THz radiation is collected and refocused on the detector using two off-axis parabolic (OAP) mirror with a diameter of 50.8 mm and focal lengths of 50.8 mm and 101.6 mm, respectively. In order to fully characterize the THz radiation, three different methods are implemented: a calibrated pyroelectric power meter (THz 20, SLT GmbH), a standard electro-optic sampling setup or a sensitive THz camera (RIGI Camera, Swiss Terahertz). To filter out the residual laser radiation and the generated green light after BNA, Polytetrafluoroethylene (PTFE) sheets (with 89% averaged THz transmission) and black pieces of cloth (with 48% averaged THz transmission) are used.

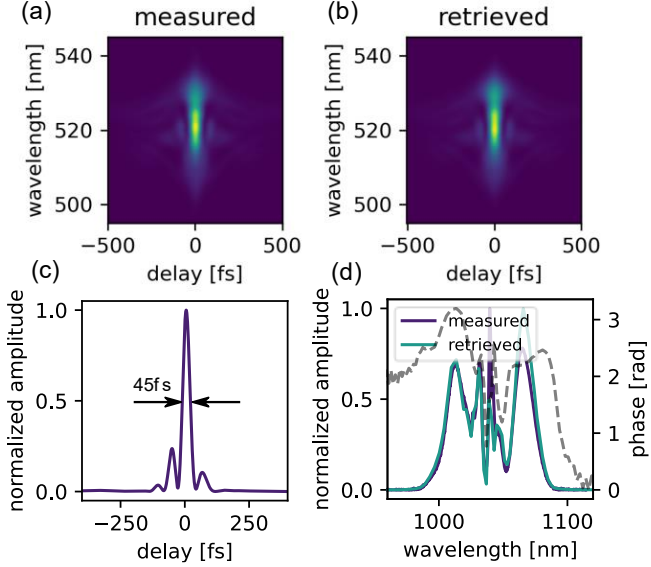

Fig. 2. Compressed pulse characterization: (a) measured FROG trace. (b) retrieved FROG trace. (c) retrieved electric field shows transform limited pulse width of about 45 fs. (d) retrieved spectrum and measured spectrum with OSA. The gray dashed line indicates the spectral phase after retraval.

## III. Results

In the first measurement set, the power meter is placed at the focus of the second OAP to measure the THz average power. The power meter is calibrated and optimized at a modulation frequency of 18 Hz at the German metrology institute (Physikalisch-Technische Bundesanstalt, PTB). Therefore, the frequency of the chopper before BNA is set to 18 Hz. In Fig. 3, the left axis shows the THz power versus pump power on the crystal and the right axis indicates THz conversion efficiency. We can pump the crystal up to 4.7 W without any irreversible damage on the crystal. This maximum pump power corresponds to an average intensity of 470 W/cm$^2$ and the peak intensity of 17 GW/cm$^2$. We reach a maximum THz average power of 5.6 mW. The calculated efficiency at the maximum THz power is 0.12%., which is in the same order of magnitude than previously obtained 0.2% at a repetition rate of 10 Hz [16] and 0.8% at 1 kHz in[17].

In order to detect the THz electric field, the power meter is replaced with a 0.2 mm GaP detection crystal in the EOS setup to sample the generated THz trace using ~200 mW of laser power. The data is acquired using a lock-in amplifier which records the signal out of the balanced photodetector and the digitized position of the shaker. The modulation frequency of the pump beam is used as a reference for the lock-in amplifier, and it is set to 2.6 kHz. The bandwidth of 300 Hz is chosen for the low-pass filter of the lock-in amplifier and the frequency of the shaker to sample the THz trace is set to 0.5 Hz. Fig. 4(a) shows the THz trace in the time domain averaged over 140 traces and recorded in 70 s in unpurged conditions. The corresponding power spectrum on a logarithmic scale is obtained by Fourier transform from the measured THz trace and is shown in Fig. 4(b). The spectrum has a wide bandwidth which spans up to about 7.5 THz with a high dynamic range of about 75 dB. The smooth and dense wideband spectrum is facilitated by favorable phase-matching conditions at 1035 nm driving wavelength.

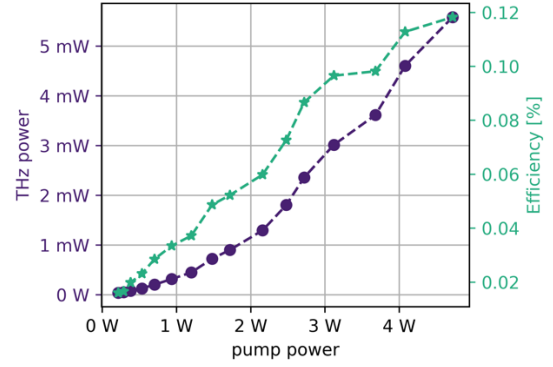

Fig. 3. Left axis, purple dashed line with circles: THz power vs. pump power, right axis, turquoise dashed line with star: THz conversion efficiency vs. pump power.

To verify the actual generated spectral bandwidth in BNA, we numerically model the THz generation process by solving the coupled wave equations in 1+1D, considering the temporal dimension and the propagation direction. The simulation considers phase matching, pump depletion, and the nonlinear susceptibility of this material [18]. The refractive index and absorption coefficient of BNA in the THz regime are taken from [19] and the nonlinear susceptibility is taken from [5]. The blue shaded area in Fig. 4(b) represents the simulated spectrum as generated in 0.65 mm BNA, showing excellent agreement with the measured spectrum in frequencies below 4 THz. Above 4 THz, the difference is due to various low-pass filtering effects in the EOS detection i.e., the response function of GaP, the increasing absorption of PTFE, and the low-pass filtering of the lock-in amplifier in combination with the shaker.

In view of future applications of this system in spectroscopy, we provide here an estimate for the THz peak electric field reached in our setup at the maximum average power of 5.6 mW using the approximation of a Gaussian THz beam and by measuring the spot size with a bolometric THz camera [20]. It is, however, to be noted that this method is only an estimated value given various uncertainties in the measurement [21], particularly for such broadband beams. In this regard, the power meter is replaced by the camera to measure the THz spot size. To characterize the THz pulse, the Gaussian envelope of the main peak of the trace shown in Fig. 4(a) is used. The intensity of the main peak can be calculated as follows:

$$I_{THz} = 0.94 \frac{W_{THz}}{A_{eff}\tau_{THz}} \quad (1)$$

Where, $A_{eff}$ is the effective area calculated using the diameter at $1/e^2$ level in the focal plane of the second parabolic mirror and $\tau_{THz}$ is the THz pulse duration at full width half maximum (FWHM). For a multi-period trace which we have from the EOS data (see Fig. 4(a)), the peak intensity ($I_{THz}$ in the formula) should be estimated for the main half-period (see dashed gaussian fit in Fig. 5(a)). Moreover, $W_{THz}$ indicates the effective energy of this main peak. The calculated half period temporal width is about 0.13 ps. Fig. 5(b) shows the THz spot in the focus of the second OAP mirror with $1/e^2$ diameter of 2.65 mm×2.25 mm.

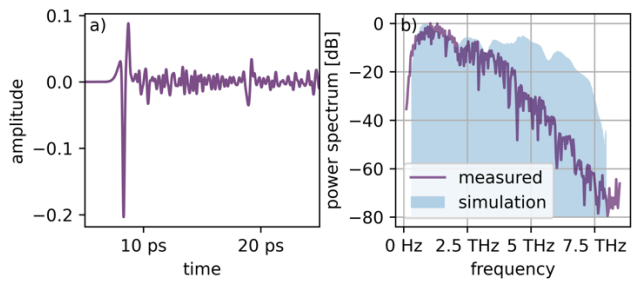

Fig. 4. Electro-optic sampling: (a) THz trace in time domain averaged over 140 traces in 70 s of measurement time. (b) corresponding spectrum in purple line and the simulation result as blue shaded area.

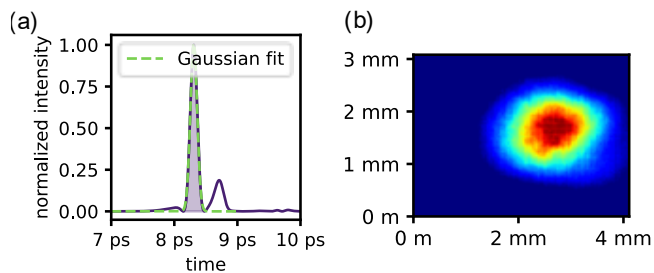

Fig. 5. THz peak electric field estimation: (a) normalized THz intensity profile with a gaussian fit on the main half period. (b) THz spot image in the focus of second OAP with the full width half maximum dimension of 1.33 mm×1.56 mm.

The THz Electric field strength is then given by:

$$E_{THz} = \sqrt{\frac{I_{THz}}{\varepsilon_0 c}} \qquad (2)$$

yielding a THz peak electric field of 29 kV/cm.

## IV. Conclusion and outlook

In conclusion, we show a high average power, broadband and high dynamic range THz-TDS pumped with a 1035 nm laser at 540 kHz repetition rate with a pulse duration of 45 fs. Using a driving power of 4.7 W, the maximum measured THz power is 5.6 mW, which is the highest THz power obtained using BNA to the best of our knowledge. The maximum calculated conversion efficiency is 0.12%, which is comparable to values obtained at much lower repetition rates using this crystal. This source represents a unique tool for a variety of time-resolved THz spectroscopy experiments, currently limited by dynamic range and bandwidth, potentially for nonlinear THz spectroscopy. We believe further THz power upscaling is possible well into tens of mW by further optimizing the cooling of the crystal, operating in purged conditions and finely tuning focusing conditions and beam chopping duty cycle.

## Acknowledgments

This work was funded by the Deutsche Forschungsgemeinschaft (DFG, German Research Foundation) – Project-ID 287022738 – TRR 196 MARIE, and in part by the Alexander von Humboldt Stiftung (Sofja Kovalevskaja Preis). It was further funded by the Deutsche Forschungsgemeinschaft (DFG, German Research Foundation) under Germany's Excellence Strategy – EXC-2033 – Projektnummer 390677874 - RESOLV. Additionally, we acknowledge support by the DFG Open Access Publication Funds of the Ruhr-Universität Bochum and by the MERCUR Kooperation project "Towards an UA Ruhr ultrafast laser science center: tailored fs-XUV beam line for photoemission spectroscopy".

## Disclosures

The authors declare no conflicts of interest.

## Data Availability Statement

Data underlying the results presented in this paper are not publicly available at this time but may be obtained from the authors upon reasonable request.

## References

1. S. W. Jun and Y. H. Ahn, Nat Commun **13**, 3470 (2022).
2. P. N. Nguyen, H. Watanabe, Y. Tamaki, O. Ishitani, and S. Kimura, Sci Rep **9**, 11772 (2019).
3. A. A. Gowen, C. O'Sullivan, and C. P. O'Donnell, Trends in Food Science & Technology **25**, 40 (2012).
4. S. Mansourzadeh, D. Damyanov, T. Vogel, F. Wulf, R. B. Kohlhaas, Bj. Globisch, T. Schultze, M. Hoffmann, J. C. Balzer, and C. J. Saraceno, IEEE Access **9**, 6268 (2021).
5. M. Jazbinsek, U. Puc, A. Abina, and A. Zidansek, Applied Sciences **9**, 882 (2019).
6. S.-J. Kim, B. J. Kang, U. Puc, W. T. Kim, M. Jazbinsek, F. Rotermund, and O.-P. Kwon, Advanced Optical Materials **9**, 2101019 (2021).
7. U. Puc, T. Bach, P. Günter, M. Zgonik, and M. Jazbinsek, Advanced Photonics Research **2**, 2000098 (2021).
8. C. Vicario, A. V. Ovchinnikov, S. I. Ashitkov, M. B. Agranat, V. E. Fortov, and C. P. Hauri, Opt. Lett. **39**, 6632 (2014).
9. C. Vicario, M. Jazbinsek, A. V. Ovchinnikov, O. V. Chefonov, S. I. Ashitkov, M. B. Agranat, and C. P. Hauri, Opt. Express **23**, 4573 (2015).
10. C. Gollner, M. Shalaby, C. Brodeur, I. Astrauskas, R. Jutas, E. Constable, L. Bergen, A. Baltuška, and A. Pugžlys, APL Photonics **6**, 046105 (2021).
11. H. Zhao, T. Wu, Y. Tan, G. Steinfeld, Y. Zhang, C. Zhang, L. zhang, and M. Shalaby, in *2019 44th International Conference on Infrared, Millimeter, and Terahertz Waves (IRMMW-THz)* (2019), pp. 1–1.
12. C. Gollner, H. Lee, X. Jiaqi, C. Weber, E. Sollinger, V. Stummer, A. Baltuska, Y. Zhang, A. Pugzlys, and M. Shalaby, in *2020 45th International Conference on Infrared, Millimeter, and Terahertz Waves (IRMMW-THz)* (2020), pp. 1–2.
13. T. O. Buchmann, E. J. Railton Kelleher, M. Jazbinsek, B. Zhou, J.-H. Seok, O.-P. Kwon, F. Rotermund, and P. U. Jepsen, APL Photonics **5**, 106103 (2020).
14. S. Mansourzadeh, T. Vogel, M. Shalaby, F. Wulf, and C. J. Saraceno, Opt. Express **29**, 38946 (2021).
15. N. Nilforoushan, T. Apretna, C. Song, T. Boulier, J. Tignon, S. Dhillon, M. Hanna, and J. Mangeney, Opt. Express, OE **30**, 15556 (2022).

16. F. Roeder, F. Roeder, F. Roeder, M. Shalaby, B. Beleites, F. Ronneberger, A. Gopal, A. Gopal, and A. Gopal, Opt. Express, OE **28**, 36274 (2020).
17. H. Zhao, Y. Tan, T. Wu, G. Steinfeld, Y. Zhang, C. Zhang, L. Zhang, and M. Shalaby, Appl. Phys. Lett. **114**, 241101 (2019).
18. T. Hattori and K. Takeuchi, Opt. Express **15**, 8076 (2007).
19. K. Miyamoto, S. Ohno, M. Fujiwara, H. Minamide, H. Hashimoto, and H. Ito, Opt. Express, OE **17**, 14832 (2009).
20. D. S. Sitnikov, S. A. Romashevskiy, A. V. Ovchinnikov, O. V. Chefonov, A. B. Savel'ev, and M. B. Agranat, Laser Phys. Lett. **16**, 115302 (2019).
21. L. Guiramand, J. E. Nkeck, X. Ropagnol, X. Ropagnol, T. Ozaki, and F. Blanchard, Photon. Res., PRJ **10**, 340 (2022).